\documentclass{aa}
\usepackage{psfig}
\usepackage{amssymb}
\def\P{PSR~B1937+21}
\def\B{{\it Beppo}SAX}
\def\nh{$N_{\rm H}$}
\newcommand{\fu}{erg~cm$^{-2}$~s$^{-1}$}
\newcommand{\HI}{H\,{\sc i}}
\newcommand{\HII}{H\,{\sc ii}}
\def\fdeg{\hbox{$\,.\!\!^{\circ}$}}
\def\loe{\lower 0.6ex\hbox{${}\stackrel{<}{\sim}{}$}}
\def\goe{\lower 0.6ex\hbox{${}\stackrel{>}{\sim}{}$}}

\begin{document}
\title{\B\ observation of PSR B1937+21}
\author{L. Nicastro\inst{1}
 \and G. Cusumano\inst{1}
 \and O. L\"ohmer\inst{2}
 \and M. Kramer\inst{3}
 \and L. Kuiper\inst{4}
 \and W. Hermsen\inst{4}
 \and T. Mineo\inst{1}
 \and W. Becker\inst{5}
}
\institute{IASF--CNR, Via U. La Malfa 153, 90146 Palermo, Italy
\and Max--Planck--Institut f\"ur Radioastronomie, Auf dem H\"ugel 69,  
 53121 Bonn, Germany
\and  University of Manchester, Jodrell Bank, Macclesfield SK11 9DL,
 United Kingdom
\and  SRON National Institute for Space Research, Sorbonnelaan 2, 3584 CA
 Utrecht, The Netherlands
\and Max--Planck--Institut f\"ur extraterrestrische Physik,
 Giessenbachstra{\ss}e, 85740 Garching, Germany
}

\date{Received 27 June 2003 / Accepted 10 October 2003}

\abstract{
We present the results of a \B\ observation of the fastest
 rotating pulsar known:
 \P.  The $\sim 200$ ks observation (78.5 ks MECS/34 ks LECS on-source time)
 allowed us to investigate with high
 statistical significance both the spectral properties and the pulse
 profile shape.
 The pulse profile is clearly double peaked at energies $\gtrsim 4$ keV.
 Peak widths are compatible with the instrumental time resolution
 and the second pulse lags the main pulse 0.52 in phase,
 like is the case in the radio.
 In the 1.3--4 keV band we detect a $\sim 45\%$ DC component;
 conversely the 4--10 keV pulsed fraction is consistent with 100\%.
 The on-pulse spectrum is fitted with an absorbed power-law of spectral index
 $\sim 1.2$, harder than that of the total flux which is $\sim 1.9$.
 The total unabsorbed
 (2--10 keV) flux is $F_{2-10} = 4.1\times 10^{-13}$ \fu, implying a
 luminosity of $L_X = 5.0\times 10^{31} \, \Theta$ ($d/3.6$ kpc)$^2$
 erg s$^{-1}$ and a X-ray efficiency of $\eta = 4.5\times 10^{-5}\, \Theta$,
 where $\Theta$ is the solid angle spanned by the emission beam.
 These results are in agreement with those obtained by ASCA
 and a more recent Rossi-XTE observation.
 The hydrogen column density $N_{\rm H} \sim 2\times 10^{22}$ cm$^{-2}$
 is $\sim 10$ times higher than expected from the radio dispersion measure
 and average Galactic density of e$^-$.
 Though it is compatible (within $2 \sigma$) with
 the Galactic (\HI\ derived) value of $\sim 1\times 10^{22}$ cm$^{-2}$,
 inspection of dust extinction maps reveal that the pulsar falls
 in a highly absorbed region. In addition, 1.4 GHz radio map shows that
 the nearby
 (likely unrelated) \HII\ source 4C21.53W is part of a circular emission
 region $\sim 4'$ across.
}

\maketitle

\section{Introduction}

The various X-ray satellites, from ROSAT to CHANDRA, observed about a dozen
 millisecond pulsars (MSPs) (see e.g. Becker 2001, Becker \& Aschenbach 2002).
 Such observations demonstrate that the X-ray emission from MSPs is
 mainly not of
 thermal origin and those with the hardest spectra appear to be
 objects with strong magnetic fields $B_L$ at the light-cylinder
 radius $R_L$ ($R_L = cP/2\pi$ where $P$ is the spin period of the MSP;
 see Saito et al. 1997,
 Kuiper et al. 1998, Kuiper et al. 2000, Becker 2001).
 However results, in the soft ``ROSAT energy band'' 0.1--2.4 keV
 from a Chandra observation of 47 Tuc,
 suggest that significant thermal emission (at energies $kT\simeq 0.2$ keV)
 can be produced from MSPs (see Grindlay et al. 2002
 for details). This appears not to be the case for PSR B1821$-$24 (see e.g.
 Becker et al. 2003).
 Correlation between spin-down energy loss and X-ray luminosity
 was investigated by several authors
 (e.g. Verbunt et al. 1996, Becker \& Tr\"umper 1997,
 Takahashi et al. 2001, Possenti et al. 2002),
 suggesting that the claimed law $L_X$(2--10 keV) $\propto\dot{E}^{\gamma}$,
 with $\gamma$ in the range 1.0--1.5, is
 valid for MSPs and ``ordinary'' pulsars in the same way.

 Unfortunately, good spectral and temporal informations exist only
 for about half of the targeted sources.
 In fact (with the exception of the data collected by Rossi-XTE and, partially,
 by ASCA and BeppoSAX)
 MSP X-ray observations are usually affected not only by low
 statistics but sometimes also by insufficient time accuracy to perform
 detailed periodicity and timing analyses.

\P, with a period $P\simeq 1.56$ ms, is the first and still the fastest
 rotating MSP known. Its radio pulse profile is double peaked with phase
 separation of about 0.52.
 In spite of its low surface magnetic field strength of
 $B_S = 3.2\times 10^{19} (P \dot{P})^{1/2} = 4.1\times 10^8$ G,
 its magnetic field at the light-cylinder
 is the highest of all known pulsars: $B_L = B_S (R_S/R_L)^3 = 2.97\times 10^8
\dot{P}^{1/2} P^{-5/2}
 \simeq 1\times 10^6$ G (it is $R_L\simeq 7.4 \times 10^6$
 and the neutron star radius $R_S = 10^6$ cm), very similar to that of
 the Crab pulsar. With a spin-down luminosity of $\dot{E} = 4\pi^2 I \dot{P}/P^3
 \simeq 1\times 10^{36}$ erg s$^{-1}$ ($I=10^{45}$ g cm$^2$,
 momentum of inertia), the derived spin-down flux density
 $\dot{E}/(4\pi (d/3.6\; {\rm kpc})^2)\simeq 7\times 10^{-10}$ \fu\
 is similar to that of PSR B1821$-$24 but it is 10 times
 larger than that of PSR J0218+4232,
 i.e. the other two MSPs detected in X-rays with good statistics
 which have high luminosities and show non-thermal emission only.

 Analysis of a sensitive observation performed by Rossi-XTE has revealed
 that the X-ray peaks (see below) are almost perfectly aligned with
 the radio giant pulses (Cusumano et al. 2003).
Such giant pulses are short bursts of emission with a flux density
exceeding that of the whole profile by a factor of 10 to 100 or
more. Having been discovered first for the Crab pulsar (Staelin \&
Reifenstein 1968), they have also been detected for PSR B1937+21
(Wolszczan et al.~1984, Backer 1995), and more recently for PSR
B1821$-$24 (Romani \& Johnston 2001), PSR B0540$-$69 (Johnston \&
Romani 2003) and PSR B1112+50 (Ershov \& Kuzmin 2003).
The observations suggest that the radio giant pulses
are related to the high energy emission of these pulsars, possibly
originating in outer gaps (Romani \& Johnston 2001). Interestingly,
Kinkhabwala \& Thorsett (2000) report that the giant pulses of \P\
appear in windows of $\sim 55\div70$ $\mu$s located after the main and
interpulse peaks, with a typical duration of less than 10 $\mu$s.
Thereby, the location of the X-ray peak coincides with the giant
pulses and not with the normal radio profile, strongly supporting
the idea about a common origin of giant pulses and high energy emission.

 In this paper we present the results of the temporal and spectral analysis
 of a \B\ observation of \P.
 Compared to those performed with ASCA (Takahashi et al. 2001)
 our observation exploits the better instrumental sensitivity and a
 longer exposure
 time. This allows the detection of the pulse profile interpulse and
 source photons down to energies of 0.5 keV and enables us
 to put constraints on the \nh\ toward the pulsar.
 Our results are compared with those obtained with ASCA
 and, more recently, with Rossi-XTE (Cusumano et al. 2003).

\section{Observation and spatial analysis \label{obssec}}
\B\ observed \P\ on May $1^{\rm st}$ 2001.
 Pointing coordinates were those from the radio timing.
 The $\sim 200$ ks observation time resulted in a total effective on-source
 time of $\sim 78.5$ ks and $\sim 34$ ks with the MECS (1.3--10 keV; Boella
 et al. 1997) and LECS (0.1--10 keV; Parmar et al. 1997), respectively.
 To extract the source photons from LECS and MECS we
 followed two alternative ways:
\begin{itemize}

\item
 extraction of the photons from the circular region and in the energy range
 which optimize the signal to noise ratio: they were $3'$, 0.5--8 keV for
 the LECS and $2'$, 1.3--10 keV for the MECS, resulting in 106 and 385
 collected photons, respectively.
 The local background (in a circular region in the field of view $10'$
 away from the pulsar) was compared to that obtained using archival data of
 blank sky observations. We found that the local background is
 $\sim 12\%$ higher for both LECS and MECS;
 background photons were 50 and 123, respectively.
\item a Maximum Likelihood (ML) approach to extract the number of counts
 assigned to the pulsar taking into account the presence of other sources
 and the background simultaneously (see e.g. Kuiper et al. 1998).
\end{itemize}

\section{Temporal analysis}
\begin{table}
\caption{JPL DE200 ephemeris of PSR B1937+21 and derived parameters.
 \label{tabeph}}
\begin{flushleft}
\begin{tabular}{ll}
\hline
\noalign{\smallskip}
{\bf Parameter} &  {\bf Value} \\
\hline
\noalign{\smallskip}
Right Ascension (J2000) &  19$^{\rm h}$ 39$^{\rm m}$ 38\fs5600084(8) \\
Declination (J2000) &  21$^\circ$ $34'$ 59\farcs13548(14)   \\
Frequency ($f$, Hz) &  641.928246349481(9) \\
Freq. derivative ($\dot{f}$, Hz s$^{-1}$) & $-4.330999(2)\times 10^{-14}$ \\
Freq. $2~{\rm nd}$ deriv. ($\ddot{f}$, Hz s$^{-2}$) & $1.558(8)\times 10^{-26}$ \\
DM$^{\rm a}$ (cm$^{-3}$ pc) &  71.03998(6) \\
Epoch of ephemeris (MJD) &  52328.0 \\
Range of validity (MJD) &  45986 -- 52764 \\
RA proper motion (mas yr$^{-1}$ ) &  $-0.128(8)$ \\
Dec proper motion (mas yr$^{-1}$) &  $-0.486(12)$ \\
\hline
\noalign{\smallskip}
Gal latitude ($l$) & 57.5215 \\
Gal longitude ($b$) &  $-0.2697$ \\
Parallax ($\pi$, mas) & $<0.28$  \\
Distance$^{\rm b}$ ($d$ -- from DM, kpc) & 3.6 \\
Surface magnetic field ($B_S$, G) & $4.1\times 10^8$ \\
Light-cylinder $B$ ($B_L$, G) & $1.0\times 10^6$ \\
Spin-down energy loss ($\dot{E}$, erg s$^{-1}$) & $1.1\times 10^{36}$  \\
$\dot{E}/4\pi d^2$ (erg cm$^{-2}$ s$^{-1}$) & $7.1\times 10^{-10}$ \\
\noalign{\smallskip}
\hline
\end{tabular}
\begin{list}{}{}
\item Note: obtained from observations with the Effelsberg 100-m telescope
 and Arecibo archival data.
 Figures in parentheses represent uncertainties ($2\sigma$) in the
 last digits quoted.
\item[$^{\rm a}$] Dispersion measure (DM) varies with time.
\item[$^{\rm b}$] From NE2001 model (Cordes \& Lazio 2002).
\end{list}
\end{flushleft}
\end{table}
The arrival times of the extracted photons were converted to the Solar System
Barycentric (SSB) frame using the BARYCONV\footnote{see
www.asdc.asi.it/bepposax/software/saxdas/} code and then searched
for periodicities by folding the MECS data with trial frequencies around the
radio one\footnote{
Folding of the lower statistics LECS data did not reveal any
 modulation at $> 2\sigma$ level.}.
The timing ephemeris listed in Table \ref{tabeph} was obtained from observations
with the Effelsberg 100-m radio telescope and archival Arecibo data.
Details are reported in Cusumano et al. (2003).
We found a clear signal, but the corresponding frequency $f_{X}=641.92825315$
Hz deviated from the radio value by $(f_{R}-f_{X})\simeq -7.6\times 10^{-6}$ Hz.
This difference is about ten times the statistical error
($\sim 8 \times 10^{-7}$ Hz).
It could be ascribed to incorrect photon time markers,
likely due to long term degradation of the on-board clock.
This effect is relevant for long duration observations only
(see Nicastro et al. 2002 for more details).

\begin{figure}
\centerline{\psfig{file=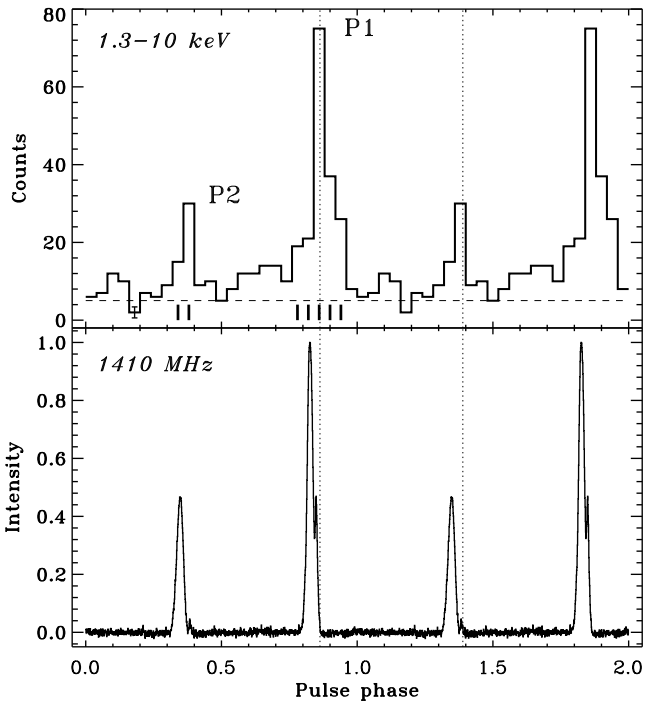,width=8.8cm,clip=} }
\caption{
 Radio pulse profile and the 1.3--10 keV \B\ X-ray profile phase aligned
 with the radio giant pulses (see text)
 marked by the two vertical dotted lines (see Cusumano et al. 2003).
 Pulse 2 lags pulse 1 in phase by 0.52, like for the radio.
 The small vertical bars under the X-ray profile mark the phase bins
 used for on-pulse spectral analysis (see text).
\label{profxr}}
\end{figure}
\begin{figure}
\centerline{\psfig{file=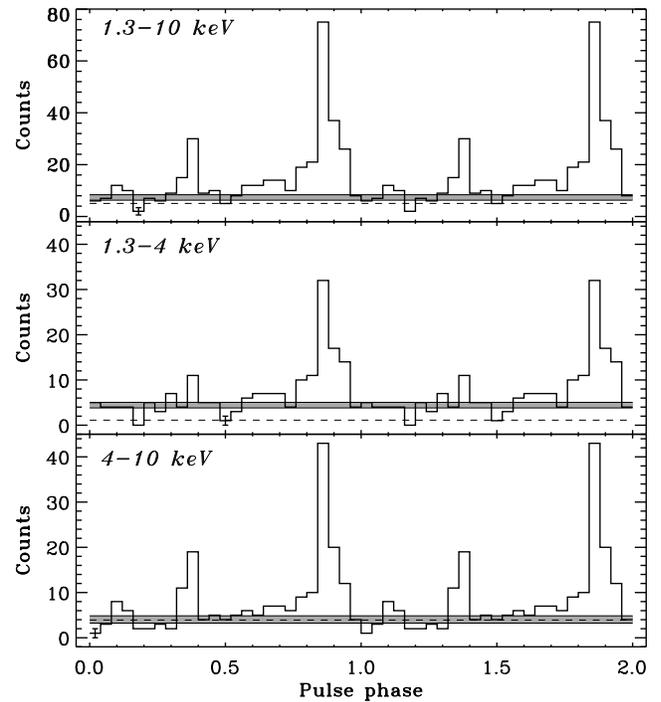,width=8.8cm,clip=} }
\caption{
 The X-ray pulse profile in the full MECS X-ray band
 1.3--10 keV and two sub-bands. The shaded areas show the estimated
 DC level ($\pm 1 \sigma$); the dashed lines indicate the measured
 background level. In the 1.3--4 keV energy band the secondary peak
 is detected at $< 2 \sigma$ level above DC level;
 in the 4--10 keV band, instead, the detection level is $>4 \sigma$
 (above background or DC level).
\label{prof3x}}
\end{figure}
Figure \ref{profxr} shows the folded light curve of the MECS data together with
 the radio one obtained with the 100-m Effelsberg radio-telescope.
 The X-ray profile is double peaked: the primary peak (P1) is detected at
 $10 \sigma$ level, the secondary peak (P2) at $\gtrsim 4 \sigma$.
 We performed a Gaussian fit on the
 two peaks. The derived phase widths are
 $\sigma_{\rm P1}=0.030$ and $\sigma_{\rm P2}=0.024$, respectively.
 This values translate in $\sim 100$ $\mu$s FWHM, fully compatible with the
 instrument time resolution limit.
 The phase separation of the two peaks (primary to secondary)
 is $0.52\pm0.02$ (90\% confidence),
 where the error is derived from the statistical uncertainties in the peak
 position gaussian fits:
 $\pm 0.007$ and $\pm 0.010$ for P1 and P2, respectively,
Figure \ref{prof3x} shows the X-ray pulse profile in 3 energy ranges.
 Non-detection of the interpulse in the ASCA data is explained by their
 lower statistical significance (227 total photons in 1.7--6.5 keV).

To quantify the pulsar DC component,
 we applied the bootstrap method proposed by Swanepoel et al. (1996),
 which allows us to estimate the off-pulse level (sky background plus
 unpulsed source component) in the pulse profile.
 We find that over the total
 energy range 1.3--10 keV the source has a pulsed fraction\footnote{It is
 $f=$ the percentage of total counts above DC level with respect to the
 net total counts in the
 full profile above the background level (determined in the spatial analysis).}
 of $f=85\pm 5\%$ reducing to $54\pm 7\%$ at energies 1.3--4 keV.
 In the 4--10 keV band the detected photons
 are $\sim 100\%$ pulsed.
 Changing the folding initial phase causes the pulse profile to change
 slightly but the detection level of the two peaks and the
 pulsed fraction remain well within the statistical error.
 The ASCA pulsed fraction of 59\% is compatible with our results.

There is an indication for an increase of the strength of P2
 compared to P1 with energy. Using the bins marked in Fig. \ref{profxr}
 as on-pulse phases, $P1/P2 = 5.13\pm 1.69$ for the band 1.3--4 keV
 and $P1/P2 = 2.99\pm 0.84$ for the band 4--10 keV.

 A joint analysis of radio data (Jodrell Bank) and X-ray data
 (ASCA) led Takahashi et al. (2001)
 to claim an alignment of the main X-ray peak with the radio interpulse.
 The \B\ on-board clock absolute accuracy does not allow us to align
 our profile to the radio one. However we noticed earlier
 (Nicastro et al. 2002) that the
 radio and X-ray peaks have both a phase separation of $\simeq 0.52$
 (primary to secondary),
 then leading us to the conclusion that their
 result is likely incorrect.
 Results from a Rossi-XTE observation has confirmed
 this doubt showing that the X-ray peaks are almost perfectly aligned with
 the radio giant pulses (see Cusumano et al. 2003).

\section{Spectral analysis}
\begin{table}
\caption{Power-law fit parameters.
\label{spetab}}
\begin{tabular}{lcccc}
\hline
\noalign{\smallskip}
 Spectrum & $K^{\rm a}$ & $\alpha$ &
 $N_{\rm H}$ & $F^{\rm b}_{2-10}$ \\
 &  &  &  ($10^{22}$ cm$^{-2}$) & \\
\hline
\noalign{\smallskip}
Total & $1.5^{+0.2}_{-0.2}$ & $1.94^{+0.13}_{-0.11}$ &
  $2.16^{+0.90}_{-0.65}$ & $4.1\pm 0.6$  \\
\noalign{\smallskip}
Pulsed$^{\rm c}$ & 0.33 & $1.21^{+0.15}_{-0.13}$ &
  $2.1^{+1.9}_{-1.1}$ & $3.0\pm 0.5$  \\
\noalign{\smallskip}
\hline
\end{tabular}
\begin{list}{}{}
\item Note: all quoted uncertainties are $2\sigma$ (95\%) confidence.
 $\chi^2_n=0.96$ for both the Total and Pulsed spectra.
 Photons distribution law $N_{ph}(E) = K\times E^{-\alpha}$.
\item[$^{\rm a}$] Normalization at 1 keV ($10^{-4}$ ph s$^{-1}$ cm$^{-2}$).
\item[$^{\rm b}$] Unabsorbed flux in 2--10 keV ($10^{-13}$ \fu).
\item[$^{\rm c}$] MECS 1.3--10 keV (see marked bins in Fig. \ref{profxr}).
\end{list}
\end{table}

In spite of the lower exposure time and sensitivity of LECS we were able
 to detect a significant ``unpulsed'' signal in the 0.5--8 keV range for
 this instrument.
 Inclusion of these softer photons in the spectral fit allowed us to better
 constrain the \nh\ toward the pulsar.
 In grouping the LECS and MECS data we applied the condition to have at least
 20 and 30 counts per bin, respectively.
 The best fitting model (in the detection range 0.5--10 keV)
 was an absorbed power-law.
 A Black Body model did not fit the total flux data satisfactorily
 ($kT\simeq 1.2$ keV and $\chi^2_n=1.5$, 13 d.o.f.).
 We then tried to fit the data with a broken power-law and a
 power-law + Black Body.
 For the latter case we also tried fixing the value of kT to various
 values in the range $0.1\div 0.5$ keV as higher values are unlikely for
 an MSP (Zavlin et al. (2002) found $kT\simeq 0.2$ keV for PSR J0437$-$4715
 and Grindlay et al. (2002) report $kT\simeq 0.22$ keV for the MSPs in 47 Tuc).
 The F-test gives chance improvement probabilities of 90\% for the
 broken power-law model and values around 50\% for the power-law + Black Body,
 then we can reject these models.
 In the energy bands 0.1--2.4 and 2--10 keV we can estimate
 \emph{absorbed} flux upper limits
 of $\sim 1\times 10^{-14}$ and $\sim 5\times 10^{-14}$ \fu\
 for a Black Body component with $kT = 0.5$ keV.
 On-pulse analysis, subtracting the off-pulse counts, was also performed;
 phase bins used to extract the {\em pulsed} photons are marked in
 Fig. \ref{profxr}.
 Fit of the 1.3--10 keV MECS data alone gave
 $N_{\rm H}=(1.2\pm0.9)\times 10^{22}$ cm$^{-2}$ and $\alpha=1.7\pm0.4$
 ($\chi^2_n=1.06$).

\begin{figure}
\centerline{\psfig{file=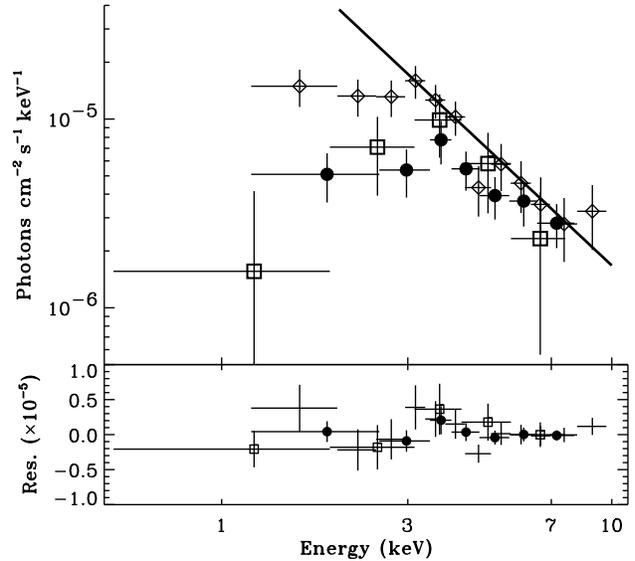,width=8.8cm,clip=} }
\caption{
 The 1.3--10 keV MECS (open diamonds) plus 0.5--8 keV LECS (open squares)
 {\em total} spectrum together with the MECS {\em pulsed} spectrum
 (filled circles).
 The straight line is the \emph{unabsorbed} power-law with $\alpha = 1.94$.
 The strong absorption below $\sim 3$ keV is evident. Fit residuals are also
 shown.
\label{speph}}
\end{figure}

In Table \ref{spetab} the best fit results
(for the absorbed power-law model) are reported; in Fig. \ref{speph}
 the total (LECS+MECS) and pulsed spectra are shown together with
 the unabsorbed power-law and the fit residuals for both the
 total and the pulsed spectra.
 We also fitted the photons collected in the off-pulse region (DC level)
 which gave, though at lower statistics,
 a consistent \nh\ and a photon index of $\sim 3.3\pm 0.5$.
 We note that the spectral index of 1.2 of the pulsed emission is harder than 
 that obtained from the total emission.
 This is a typical value for magnetospheric emission and is fully consistent
 with the value obtained by Cusumano et al. (2003) using Rossi-XTE data.
 The total unabsorbed and absorbed fluxes of \P\ in the 2--10 keV range are
 $4.1\times 10^{-13}$ and  $3.5\times 10^{-13}$ \fu, respectively.
The derived (unabsorbed) luminosity is $L_X = 5.0\times 10^{31} \, \Theta$
 ($d/3.6$ kpc)$^2$ erg s$^{-1}$ while the X-ray conversion efficiency is
 $\eta = L_X/\dot{E} = 4.5\times 10^{-5}\, \Theta$,
 where $\Theta$ is the solid angle spanned by the emission beam.
 This result is confirmed by the Rossi-XTE 2.5--17 keV spectrum
 (Cusumano et al. 2003).
 We also note that the fit parameters are consistent with the non-detection
 (Verbunt et al. 1996) in the ROSAT HRI energy range (0.1--2.4 keV).
%

\section{Discussion}

\subsection{Neutral hydrogen column density}
 Galactic \nh\ can be estimated from radio/IR measurements in three
 different ways:
 1. from the DM by assuming 10 H atoms for each e$^-$; 2. from \HI\
  measurements, assuming there is no hydrogen at molecular state; 3. from
 dust extinction using the relation $N_{\rm H}=1.79 \times 10^{21} A_V$
 of Predehl \& Schmitt (1995). When there are no direct pulsar measurements or
 three dimensional distribution models,
 points 2. and 3. make use of models of the
 integral content of H along a particular line of sight.
 They then only give upper limits for the atomic and total hydrogen content,
 respectively.

 For PSR B1821$-$24 placed in the globular cluster M28, at a
 distance of $\sim 5.5$ kpc and galactic latitude of $-5\fdeg6$,
 one finds $N_{\rm H}$
 $\simeq 1.6\times 10^{21}$ cm$^{-2}$ which is consistent with the
 estimates from Galactic \HI, dust and pulsar dispersion measure
 (Becker et al. 2003).
 Similarly consistent results are obtained for PSR J0218+4232.
 However, because of variable e$^-$/$N_{\rm H}$ ratios in the various
 Galactic directions
 and the presence of more o less dense gas clouds along the line of
 sight, discrepancies would not be surprising.

 The column density we derive from the power-law fit is 10 times greater
 than the radio DM derived one and is only
 marginally consistent (within $2 \sigma$) with the \HI\ derived Galactic value
 $1.2\times 10^{22}$ cm$^{-2}$ (Dickey \& Lockman 1990). In addition, it is
 {\em apparently} totally inconsistent with the $7.4 \times 10^{22}$ cm$^{-2}$
 obtained from the Schlegel, Finkbeiner \& Davis
 (1998; SFD) dust extinction maps (Schlegel et al. (1998).

\begin{figure}
\centerline{\psfig{file=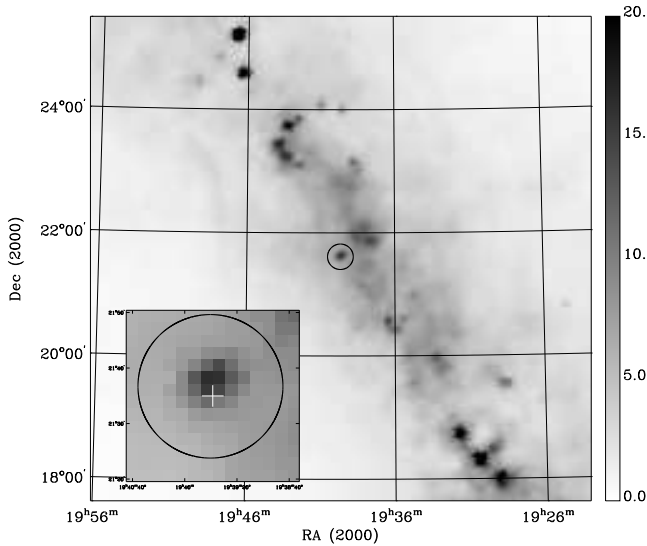,width=8.8cm,clip=} }
\caption{
 $8^\circ \times 8^\circ$ SFD (Schlegel et al. 1998) dust extinction map
 centered on \P. The Galactic plane is clearly visible.
 The circle ($24'$ diameter) is centered on IRAS 19375+2130.
 The extinction toward this source is about three times higher than toward the
 surrounding regions $\sim 20'$ away from it.
 The pulsar position is marked with a cross in the zoomed inset.
\label{dust}}
\end{figure}

 Being the line of sight to \P\ tangent to the Galactic spiral arms
 ($l \simeq 57.51$, $b \simeq -0.29$, see e.g. Fig. 3 of Watson et al. 2003),
 one can invoke a particularly low e$^-$/$N_{\rm H}$ ratio in that
 direction due to the lack of ionizing sources.
 \HI\ maps report the integral Galactic column density but
 do not account for hydrogen in molecular form and then, in particular
 for objects in the
 Galactic plane, they must generally be regarded as lower limits.
 A more reliable check of the X-ray determined
 \nh\ can come from the investigation of dust extinction maps.
 Using the afore mentioned Predehl \& Schmitt (1995) relation,
 Verbunt at al. (1996) set an upper limit of
 $2.2 \times 10^{21}$ cm$^{-2}$ for the column density (i.e. in agreement
 with the DM value). From the more recent SFD map,
 as shown in Fig. \ref{dust}, we note that \P\ is almost at the center of an
 enhanced absorption region.
 The map has a pixel resolution of $2\farcm37\times2\farcm37$ and
 spatial resolution of 6\farcm1 (FWHM).
 The enhanced region is then consistent with a point source at coordinates
RA (J2000) $\simeq 19^{\rm h}$ 39\fm5
Dec (J2000) $\simeq 21^\circ$ 37\farcm5.
 Given the mentioned spatial resolution,
 it is compatible with the position of \P. However 
 at this position there is a strong radio and IR source: IRAS 19375+2130
 ($\equiv$ 4C21.53W $\equiv$ G57$-$0.27 $\equiv$ G57.55$-$0.27).
 This source was extensively investigated since the discovery of \P\
 (see e.g. Becker \& Helfand 1983). It is listed as an ultra-compact
 \HII\ region and selected by Wood \& Churchwell (1989) as a candidate
 site of OB star formation.
 To estimate the visual Galactic extinction toward the pulsar, we used
 the map value of neighboring regions: it is E(B$-$V) $\sim 6$ and a value of
 $A_V\sim 19$ is derived, giving
 $N_{\rm H}=3.3\times 10^{22}$ cm$^{-2}$.
This is some 50\% higher than the X-ray derived \nh\ for \P.
The observed high dust/gas ratio can then be justified
without requiring a very large distance to the pulsar.
However it has to
be larger than the NE2001 model derived one (Cordes \& Lazio 2002)
as 3.6 kpc is a lower limit derived from timing (see below).
 Considering that the SFD map predictions are not very reliable at galactic
 latitudes $|b|<5^\circ$ (Schlegel et al. 1998),
 the consistency with the X-ray derived value is remarkable.

\subsection{The pulsar distance}
 Timing observations of the pulsar provide proper motion measurements
 and an upper limit on the timing parallax of $\pi < 0.28$ mas
 (see Table \ref{tabeph}).
 This information can be used to infer that the DM distance of 3.6 kpc is,
 in fact, a lower limit. An upper limit can be derived
 from the transversal velocity
 estimate given by the interstellar scintillation velocity
 $v_{\rm ISS} \simeq 50\pm 10$ km s$^{-1}$ found by Cordes (1990).
 Anyway, given the timing derived pulsar proper motion,
 the resulting limit of $d=v / (4.74\; \mu) = v / (4.74\times 0.5)=21$ kpc
 is not very
 constraining. While it has been shown that the scintillation velocity is
 indeed a good estimator for the true transversal velocity
 (Nicastro et al. 2001), for low velocities such as measured for MSPs
 (i.e. $\sim 30$ km s$^{-1}$), systematic effects can result in large errors.
 A campaign of scintillation study could lead to a more reliable $v_{\rm ISS}$
 estimate, but it is unlikely that a value lower than 30 km s$^{-1}$
 can be measured and therefore it can only be useful to set the Galactic upper
 limit distance of $\lesssim 10$ kpc.
 On the other hand, regular multi-frequency observations can help to
 remove the effects of ``interstellar weather'', which degrades the timing
 precision of PSR B1937+21 significantly (e.g. Backer \& Wong 1996).
 If these effects can be modeled, higher timing precision is obtainable in
 the future, possibly helping in setting tighter limits on the
 parallax measurements.

 Heiles et al. (1983) studied the \HI\ absorption spectra
 toward \P, 4C21.53W and another close-by source.
 By setting the Galactic tangent point to 5.4 kpc,
 a comparison of position and intensity of the lines for these
 three objects allowed them to conclude that
 4C21.53W is at $\simeq 10.7$ kpc and the pulsar at $\simeq 5$ kpc from us.
 Frail \& Weisberg (1990) revised this result using a new model for
 the rotation of the Galaxy. They adopt the tangent point distance
 as the pulsar distance lower limit: $d>4.6\pm1.9$ kpc.
 Watson et al. (2003) have recently confirmed the position of the \HII\
 region in the Perseus arm. Assuming a tangent point at a distance of 4.6 kpc,
 they derive a distance of 8.6 kpc by studying H$_2$CO absorption and
 H110$\alpha$ emission lines
 (H110$\alpha$ is the 4874.1570 MHz radio recombination line).
 However the data are inconclusive on the position of the
 intervening molecular cloud, giving either 3.6 or 5.5 kpc as possible
 distances, for the two sides of the tangent point, respectively.
 With the pulsar at a distance $>3.6$ kpc, it is possible that the cloud
 contributes significantly to the absorption of the low energy X-rays.

\begin{figure}
\centerline{\psfig{file=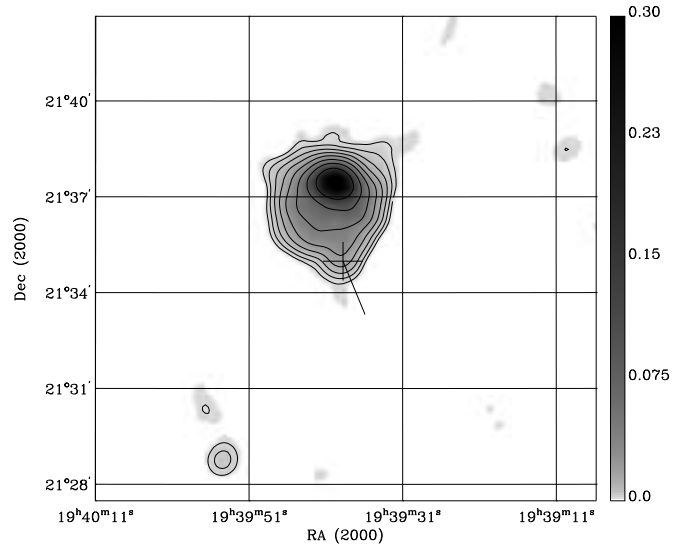,width=8.8cm,clip=} }
\caption{
 $0\fdeg25\times 0\fdeg25$ NVSS 1.4 GHz image around \P.
 The pulsar is marked with a cross with the proper motion direction toward
 South-West
 shown. The diffuse region is $\sim 4'$ across with $\sim 60$ mJy/beam at the
 center. The strong source is 4C21.53W ($\sim 360$ mJy/beam).
 Contours are log spaced.
\label{nvss}}
\end{figure}

 We also investigated the 1.4 GHz NVSS data (Condon et al. 1998).
 Figure \ref{nvss} shows the $15'\times 15'$ region around the pulsar position.
 \P\ lies on the edge of a diffuse circular region
 (flux is $\sim 50\div 60$ mJy/beam) and is moving away from it
 as indicated by the proper motion vector shown in Fig. \ref{nvss}.
 Incidentally, the vector orientation along the
 Galactic plane can explain why \P\ has the lowest z-height of all MSPs.
 The strong source to the North is the aforementioned 4C21.53W and has
 a flux density of $\sim 800$ mJy, compared to the $\sim 20$ mJy of \P.
 The non-detection of the diffuse region in the VLA maps by
 Becker \& Helfand (1983) can be explained by their use of a different array
 configuration. They used the B-array rather than the D-configuration as
 used for the NVSS, resulting in a
 much degraded sensitivity for diffuse emission.
 Though the circular shape of the region and the pulsar velocity vector could
 be suggestive of an association, the curved (externally ``compressed'')
 shape of the \HII\ region indicates a relation with the latter (see Fig. 1
 of Becker \& Helfand 1983).
A possible way to estimate the chance coincidence of \P\ with the \HII\
 ``complex'' is to select the sources of Watson et al. (2003)
 away from the inner arms, i.e. $70^\circ<l<50^\circ$ and $|b| \leq 1^\circ$,
 and assume that they all have an associated ``region'' of $4'$ diameter
 (we recall that those \HII\ regions
 were selected from the IRAS \emph{point source} catalogue,
 i.e. they have FWHM $<2'$ at 100 $\mu$m).
 The ratio of the total area of the selected 20 sources to
 $20\times2$ square degrees is $9\times10^{-4}$. We note also that in the same
 area there are 15 known pulsars, two of which are MSPs\footnote{
 http://www.atnf.csiro.au/research/pulsar/psrcat/}.
 The coincidence at this galactic longitude
 of a peculiar pulsar with such a small region
 (at a distance of 8.6 kpc the region would be $\sim 20$ pc across,
 $\sim 3.5$ pc at 3.6 kpc) remains then remarkable.

\subsection{Spectral index and magnetospheric emission}
 Table \ref{enetab} lists the main emission properties of the three MSPs
 which show non-thermal emission only. In order to account for the
 contribution to the DC flux
 by the field source listed by Becker et al. (2003), but
 unresolved by the \B\ MECS, we re-analysed the
 \B\ observation of B1821$-$24.
 The pulsed fraction was compatible with 100\%. This is consistent with the
 result by Danner et al. (1997) using ROSAT HRI data in the
 0.1--2 keV band and explains the
 similar indices of the on-pulse \B\ spectrum and the total CHANDRA
 spectrum, for which no pulse phase selection could be performed.
 As can be seen from Table \ref{enetab}, the ``total'' flux photon index of \P\
 is significantly softer than that measured for
 PSR J0218+4232 and PSR B1821$-$24.
 Also, we found that the spectrum of the pointlike DC component of PSR B1937+21
 is softer than its pulsed spectrum.
 It is interesting to note that high-resolution CHANDRA observations of PSR
 J0218+4232 (Kuiper et al. 2002) confirmed the presence of a DC component of
 this source, earlier noted in Kuiper et al. (1998) and Mineo et al. (2000).
 The spectrum of the DC component of PSR J0218+4232 appeared to be softer than
 its pulsed component, like we found here for PSR B1937+21.
 Kuiper et al. (2000) concluded that we do not expect thermal emission
 from the surface of an old, recycled MSP due to cooling, and that the
 soft DC component is most likely due to reheating of the
 polar cap area by backflowing accelerated particles.
 This has to be investigated further for \P.
 A high spatial resolution observation may reveal field sources
 and/or diffuse emission also in the \P\ field. Such information would
 allow us to better constrain both its total and pulsed emission
 properties as compared to those of PSR B1821$-$24 and PSR J0218+4232
 for which higher resolution observations already exist.
 In spite of the large $\dot{P}$ difference between \P\ and B1821$-$24,
 the X-ray production efficiencies of these two pulsars are similar; and
 similar are their $B_L$ and pulsed emission spectral indices.
 PSR J0218+4232 is a factor of 10 more efficient even though its $B_L$
 is the weakest of the three; on the other hand it has the hardest spectrum.
 We finally note that using 4.6 kpc as the distance to \P\ and as the distance
 to PSR J0218+4232 the 4 kpc upper limit recently set by Bassa et al. (2003),
 the ``pulsed'' X-ray luminosities of the three pulsars are almost identical.
 Identical are also the pulsed spectral indices if for PSR J0218+4232 we
 adopt the statistically more significant value of $\alpha=1.14$
 from Kuiper et al. (2003).

 All together these figures are not inconsistent with the claiming of a single
 $L_X-\dot{E}$ law for MSPs and ordinary pulsars suggesting that a common
 magnetospheric emission process could be responsible for most,
 if not all, of the emitted X-ray flux in the band 2--10 keV.
 It is not clear if pulsar in 47 Tuc could be peculiar in this respect given
 that they have been studied only in the soft energy range 0.1--2.4 keV
 (Grindlay et al. 2002).
\begin{table}
\caption{Energetics of \P\ compared to B1821$-$24 and J0218+4232.
\label{enetab}}
\begin{tabular}{lcccc}
\hline
\noalign{\smallskip}
 {\bf Parameter} & B1937+21$^{\rm a}$ & B1821$^{\rm b}$ & J0218$^{\rm c}$ \\
\noalign{\smallskip}
\hline
\noalign{\smallskip}
 Period ($P$, ms) & 1.558 & 3.054 & 2.323 \\
 $\dot{P}$ ($\times 10^{-19}$ s Hz) & 1.0 & 16 & 0.77 \\
 $\dot{E}$ ($\times 10^{36}$ erg s$^{-1}$) & 1.1 & 2.2 & 0.25 \\
 $B_L$ ($\times 10^6$ G) & 1.0 & 0.77 & 0.32 \\
 Distance ($d$, kpc) & 3.6 (4.6) & 5.5 & 5.7 (4.0) \\
 $\dot{E}/(4\pi d^2)^{\rm d}$ ($\times 10^{-10}$) & 7.1 (4.3) & 6.1 & 0.64 (1.3) \\
 $\alpha$ total & $1.94^{+0.13}_{-0.11} $ & $1.20^{+0.15}_{-0.13}$ &
  $0.94^{+0.22}_{-0.22}$ \\
\noalign{\smallskip}
 $\alpha$ pulsed & $1.21^{+0.15}_{-0.13} $ & $1.10^{+0.20}_{-0.20}$ &
  $0.61^{+0.32}_{-0.32}$ \\
 Pulsed frac.$^{\rm e}$ ($f$, \%) & 86 & 98 & 73 \\
 $L_{X,{\rm P}}^{\rm f}$ ($\times 10^{32}$) &
   0.37 (0.60) & 0.62 & 1.3 (0.64) \\
 $L_{X,{\rm I}}$ &
   6.4 (10) & 12 & 16 (7.9) \\
 $L_{X,{\rm E}}$ &
   5.7 & 16 & 0.67 \\
 $L_{X,{\rm P}}/\dot{E}$ ($\times 10^{-4}$) &
   0.34 (0.55) & 0.28 & 4.8 (2.4) \\
 $L_{X,{\rm I}}/\dot{E}$ ($\times 10^{-4}$) &
   5.8 (9.5) & 5.6 & 65 (32) \\
\noalign{\smallskip}
\hline

\end{tabular}
\begin{list}{}{}
\item[] Note: luminosity $L_X$ (\fu) and flux $F_X$ (erg s$^{-1}$) in 2--10 keV.
\item[$^{\rm a}$] In parenthesis figures for $d=4.6$ kpc.
\item[$^{\rm b}$] See Becker et al. (2003); Kawai \& Saito (1999).
\item[$^{\rm c}$] See Mineo et al. (2000). In parenthesis figures for $d=4.0$
 kpc (Bassa et al. 2003). Using Rossi-XTE, Kuiper et al. (2003)
 find over the range 2 to $\sim 20$ keV $1.14^{+0.03}_{-0.04}$
 for $\alpha$ pulsed.
\item[$^{\rm d}$] Spin-down flux (\fu).
\item[$^{\rm e}$] In 1.6--10 keV as derived by \B\ observations.
\item[$^{\rm f}$] Luminosity: P = pulsed ($d^2\times F_{X,P}$), I = isotropic total, 
 E = isotropic expected from the relation
 $\log L_X = 1.45 \log \dot{E} -19.5$
 (Takahashi et al. 2001).
\end{list}
\end{table}

\section{Conclusions}

We detected the double peak profile of the fastest rotating pulsar known.
The 1.3--10 keV pulsed fraction is 85\% which becomes 100\% in
the 4--10 keV band. The DC component is significantly detected only
in the low energy band 1.3--4 keV ($46 \pm 7\%$ unpulsed photons).
The pulse phase resolved spectral analysis confirms this result.
The secondary (X-ray) peak is detected at high significance above 3--4 keV
and the ratio primary/secondary decreases with energy. This suggests that
the secondary peak has a harder spectrum.
It was not possible to perform an absolute phase comparison
with the radio profile, but phase separation comparison indicates,
contrarily to the ASCA finding by Takahashi et al (2001), that
the main radio and X-ray peaks are aligned.
This indeed was confirmed by a Rossi-XTE observation which shows
that the X-ray peaks are almost perfectly aligned with the radio giant pulses
(Cusumano et al. 2003).

We measure a hard power-law spectral index $\alpha \simeq 1.2$
for the pulsed photons. This is similar to the value found for
PSR B1821$-$24 and PSR J0218+4232 (Kuiper et al. 2003).
We find a column density of $N_{\rm H} \sim 2.1\times 10^{22}$ cm$^{-2}$
toward \P. It is marginally consistent with the integral Galactic value
obtained by the Dickey \& Lockman (1990) \HI\ model
of $1.2\times 10^{22}$ cm$^{-2}$ but in agreement with the SFD map
prediction, and gives a e$^-$/$N_{\rm H}$ ratio of 1/100.

From the NVSS maps we note that the pulsar is located on the
edge of a diffuse $\sim 4'$ circular region with the strong radio/IR
source 4C21.53W and the pulsar located at opposite edges.
The pulsar also seems to move away from this region with a
transverse speed vector along the Galactic plane.
Emission/absorption line observations have shown that 4C21.53W is very likely
located in the Perseus arm at 8--9 kpc away from us (Watson et al. 2003)
whereas it is unlikely the pulsar to be too far away from the Galactic
tangent point, i.e. 4.6 kpc (Frail \& Weisberg 1990).
These results and the morphology of 4C21.53W/diffuse region suggest
they are associated, with the pulsar aligned by chance.

A detailed study of the ISM toward the pulsar could
possibly help modeling the (relatively) high timing noise of \P\ and
consequently further constrain the parallax mesurement.
However, we consider $\sim 5$ kpc to be a good distance estimate.

A high spatial resolution CHANDRA X-ray observation
can help both to verify the high \nh\ value and to study the continuum X-ray
emission properties of its close surroundings.
A better resolved 1.4 GHz radio map (obtainable from C configuration VLA
archival data) can help to study the morphology of the ``4C21.53 complex''
and to confirm the chance alignment of the diffuse emission
region with the pulsar.

\vskip 0.4cm
\begin{acknowledgements}
This research is supported by the Italian Space Agency (ASI) and
Consiglio Nazionale delle Ricerche (CNR). BeppoSAX {\em was} a major program
of ASI with participation of the Netherlands Agency for Aerospace
Programs (NIVR). Its unburned pieces sinked in the Pacific Ocean
on April 30, 2003, exactly on the date of its $7^{\rm th}$ birthday.
We would like to thank the anonymous referee whose comments and suggestions
improved the clarity of the paper.
\end{acknowledgements}

\end{document}